\documentclass[12pt,eqsecnum,floatfix,aps,prc]{revtex4}
\usepackage{epsfig}

\def\fun#1#2{\lower3.6pt\vbox{\baselineskip0pt\lineskip.9pt
  \ialign{$\mathsurround=0pt#1\hfil##\hfil$\crcr#2\crcr\sim\crcr}}}

\begin{document}
\title{Continuum-discretized coupled-channels method \\ for
four-body breakup reactions
\footnote{
A talk given at the Workshop on
Reaction Mechanisms for Rare Isotope Beams,
Michigan State University, March 9-12, 2005
(to appear in an AIP Conference Proceedings).
}}

\author{
M. Kamimura$^1$,
T. Matsumoto$^1$,
E. Hiyama$^2$,
K. Ogata$^1$,
Y. Iseri$^3$,
and M. Yahiro$^1$
}

\address{$^1$Department of Physics, Kyushu University, 
Fukuoka 812-8581, Japan}

\address{$^2$Department of Physics, Nara Women's University,
Nara 630-8506, Japan}

\address{$^3$Department of Physics, Chiba-Keizai College, 
Chiba 263-0021, Japan}

\date{\today}

\begin{abstract} 
\baselineskip=6mm
Development of the method of CDCC (Continuum-Discretized 
Coupled-Channels) from the level of three-body CDCC
to that of four-body CDCC is reviewed. 
Introduction of the pseudo-state method based on the 
Gaussian expansion method for discretizing 
the continuum states of two-body and three-body projectiles
plays an essential role in the development.
Furthermore, introduction of the complex-range Gaussian
basis functions is important to improve the CDCC for
nuclear breakup so as to accomplish that for
Coulomb and nuclear breakup.
A successful  application of the four-body CDCC to 
$^6$He+$^{12}$C scattering at 18 and 229.8 MeV is 
reported.
\end{abstract}


\maketitle
\baselineskip=6mm
\section{INTRODUCTION}
\begin{figure}[htbp]
 \includegraphics[width=0.7\textwidth,clip]{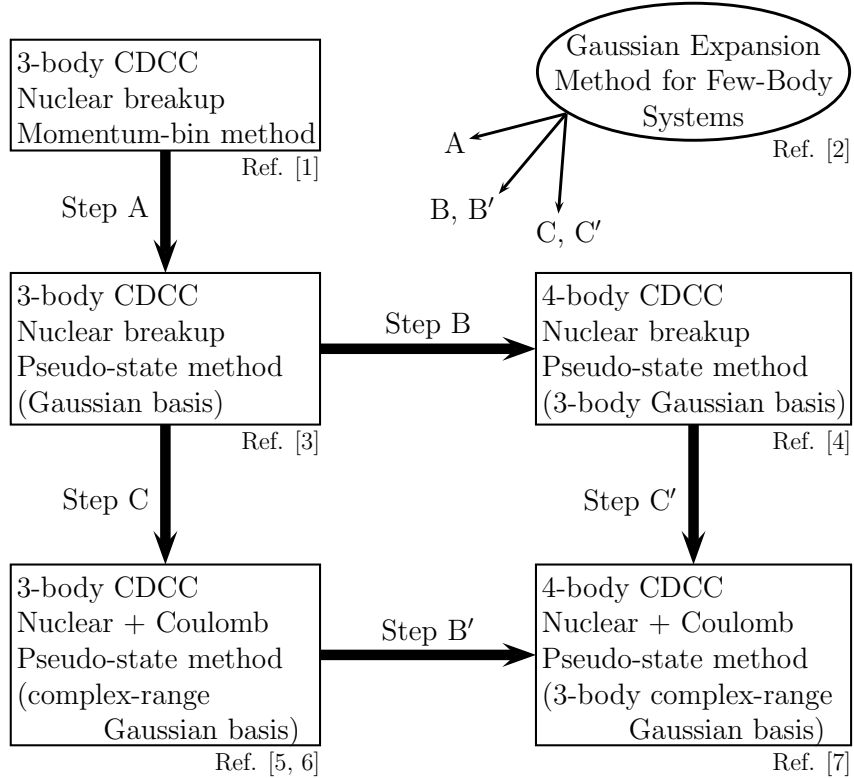}
 \caption{\baselineskip=6mm
 A flow of the improvement of the method of CDCC
 starting from the three-body CDCC for nuclear breakup
 to the four-body CDCC for Coulomb and nuclear breakup
 with the aid of the Gaussian expansion method
 for few-body systems.
 }
 \label{fig:cdcc-th}
\end{figure}

In the study of reactions induced by unstable nuclei,
analysis of the case where the projectile
is considered to be composed of three-clusters such as 
$^6$He and $^{11}$Li becomes quite important.
For this purpose, along the diagram in Fig.~\ref{fig:cdcc-th},
we have developed the three-body CDCC
(Continuum-Discretized Coupled-Channels)
for nuclear breakup of 
two-body projectiles \cite{cdcc}
into the four-body CDCC for Coulomb and nuclear breakup
of three-body projectiles.

The momentum-bin method to discretize the 
continuum states of the two-body projectiles
(such as $^6{\rm Li}=\alpha+d,\;  ^8\!{\rm B}+p$, etc.)
is not practically available to the case of three-body
projectiles.  On the basis of the Gaussian expansion method
(GEM)\cite{GEM}, we proposed, in Ref.\cite{ps-cdcc},
the pseudo-state (PS) method
to discretize the continuum states and
examined it in the case of two-body projectiles
(three-body CDCC); this is Step A in Fig.\ref{fig:cdcc-th}.
In the PS method we diagonalized the two-body Hamiltonian
of the internal motion of the projectile
using the Gaussian basis functions \cite{GEM}
and obtained  dense distribution of 
the pseudo-states, namely discretized continuum states.
An advantage of this method is that it can easily be extended
to the case of three-body projectiles by using
the GEM.
Another advantage of the
PS method is that we can derive continuous 
$S$-matrix elements as a smooth function of
the momentum of the projectile breakup states.
We found \cite{ps-cdcc}
that the $S$-matrix elements obtained by the PS method
agrees well with the  $S$-matrix elements
by the momentum-bin method with very precise bins.

As Step B in Fig.\ref{fig:cdcc-th}, we extended the
three-body CDCC (for nuclear breakup) 
to the four-body CDCC (for nuclear breakup)
using the three-body Gaussian basis functions
of GEM to obtain bound and 
pseudo-states of the three-body projectiles
\cite{fb-cdcc}.
The GEM is very suitable for describing 
bound and pseudo-states of three- and four-body systems;
it is extensively reviewed in Ref.\cite{GEM}.
The four-body CDCC was applied to the $^6$He+$^{12}$C at 
18 and 229.8 MeV.  The differential cross sections of the
elastic scattering  were well reproduced by using
the double-folding CC potentials.   

In Step C, we improved the three-body CDCC for nuclear breakup
to that for Coulomb and nuclear breakup
\cite{psc-cdcc} by using the PS method  with the complex-range
Gaussian basis functions \cite{GEM}
instead of the (real-range) Gaussian
basis functions adopted in the previous Steps.
Due to the long-ranged Coulomb coupling-potentials, 
the modelspace
required for CDCC is very large. Particularly, one must prepare
the internal wave functions of the projectile, both in bound and
continuum states, for a wide range of the internal coordinate,
say 0--100 fm, which is in general difficult for PS methods.
This can easily be achieved by using the complex-range Gaussian
basis in the case of two-body projectile.

In order to treat both Coulomb and nuclear breakup processes
at {\it intermediate energies}
with high accuracy and computational speed,
a new method was proposed in Ref.~\cite{ecdcc};
namely, a hybrid calculation with
the three-body CDCC method and the eikonal-CDCC (E-CDCC) 
method. 
E-CDCC describes the center-of-mass motion of the
projectile relative to the target by straight-line 
approximation (or by using Coulomb wave functions instead of
plane waves) and treats the excitation of the  projectile
explicitly by CDCC with the momentum-bin method or
the PS method.
E-CDCC drastically reduces computation time 
and eliminates many problems concerned with huge angular 
momentum in solving coupled-channel
equations. Thus, the hybrid calculation is expected to be 
opening the door to the systematic
analysis of Coulomb (plus nuclear) dissociation
of projectiles in the wide range of beam energies.

Finally, by Step B' (or by Step C')
we can reach the four-body CDCC for Coulomb and
nuclear breakup.  This step was not reported in the time
of the RIA workshop but was recently accomplished and
successfully applied to the $^6$He+$^{209}$Bi scattering
at 19.0 and 22.5 MeV \cite{fbc-cdcc}.

\section {METHOD OF PSEUDO-STATE CDCC FOR TWO-BODY PROJECTILES}

In the method of CDCC, the total wave function of the
scattering state $\Psi_{JM}$ is expanded in terms of 
a finite number of internal wave functions $\Psi_{nIm}(\xi)$
of the projectile:
\begin{equation}
\Psi^{JM}(\xi,{\bf R})
=\sum_{nI,L}
 [\Phi_{nI}(\xi)
\otimes \chi_{nI,L}^J ({\bf R}) ]_{JM},
\label{expansion}
\end{equation}
where ${\bf R}$ is the coordinate of
the center-of-mass of the projectile
relative to the target, and $\xi$ is the internal
coordinates of the projectile.
$I$ is the total spin of the projectile
and $n$ stands for the $n$th eigenstate. 
$\chi_{nI,L}^J$ represents the relative motion between 
the projectile and the target;
$L$ is the orbital angular momentum regarding ${\bf R}$.
The unknown function $\chi_{nI,L}^J ({\bf R}) $
are solved using the usual framework of
the coupled-channel method for discrete excited states.

The projectile internal wave functions
$\Phi_{nI}(\xi)$ include both bound states 
and discretized continuum states.
To calculate the wave functions of the latter states
the momentum-bin method has widely been utilized 
in the usual three-body CDCC calculations.
In the method the exact 
scattering wave functions are averaged within each narrow
intervals of momentum between the two constituents 
in the projectile.
But, this method is not practically
suitable for discretizing the breakup states of
the three-body projectile.

  In the pseudo-state (PS) method~\cite{cdcc,PS1, PS2},
 on the other hand, wave functions of
the discretized breakup states are
obtained by diagonalizing the internal Hamiltonian of
the projectile, which describes the relative motion of
the two constituents,
using $L^2$-type basis functions.
Since the wave functions of
such pseudo breakup states have wrong
asymptotic forms,
the PS method was mainly used in the past to describe
virtual breakup processes in the
intermediate stage of elastic scattering~\cite{PS2}
and ($d,p$) reactions~\cite{cdcc}.

In the work of Ref.\cite{ps-cdcc}, however,
we proposed the new method of pseudo-state (PS)
discretization for two-body projectiles.
It can be used not only for virtual breakup processes
in elastic scattering but also for breakup reactions.
In order to diagonalize the Hamiltonian of the two-body projectile,
we employed two types of basis functions. 
One is the conventional real-range Gaussian functions
\begin{equation}
\label{eq:RG}
\phi_{j\ell}(r)=r^{\ell}\exp\left[-(r/a_j)^2\right],
\qquad (j = 1 \mbox{--}n)
\end{equation}
where $\{a_{j}\}$ are
assumed to increase in a geometric
progression~\cite{Kamimura88,GEM}:
\begin{equation}
 a_{j}=a_1 (a_{n}/a_1)^{(j-1)/(n-1)}.
\label{aj}
\end{equation}
The other is an extension of (\ref{eq:RG})
introduced in Ref.~\cite{GEM}, i.e., the following
pairs of functions:
\begin{eqnarray}
\phi^{\rm C}_{j\ell}(r)
&=&
r^{\ell}\exp\left[-(r/a_{j})^2\right]
\cos\left[\,b\,(r/a_{j})^2\right],
\nonumber \\
\phi^{\rm S}_{j\ell}(r)
&=&
r^{\ell}\exp\left[-(r/a_{j})^2\right]
\sin\left[\,b\,(r/a_{j})^2\right],
\quad (j = 1 \mbox{--} n).
\label{eq:comp-g}
\end{eqnarray}
Here, $b$ is a free parameter, in principle, but
numerical test showed that $b =\pi/2$ is recommendable.
Both $\phi^{\rm C}_{j\ell}$ and $\phi^{\rm S}_{j\ell}$
are to be used simultaneously; 
the total number of basis is thus $2n$.
The basis functions (\ref{eq:comp-g})
can also be expressed as
\begin{eqnarray}
\phi^{\rm C}_{j\ell}(r)
&=&
\{\psi_{j\ell}^{*}(r)+\psi_{j\ell}(r)\}/2,
\nonumber \\
\phi^{\rm S}_{j\ell}(r)
&=&
\{\psi_{j\ell}^{*}(r)-\psi_{j\ell}(r)\}/(2i),
\end{eqnarray}
with
\begin{equation}
\label{CG}
\psi_{j\ell}(r) =r^{\ell}\exp[-\eta_{j} r^2],
\;\;\;
\eta_{j}=(1+i\,b)/a_{j}^2,
\end{equation}
i.e., Gaussian functions with a complex-range parameter.
We thus refer to the basis
$\phi^{\rm C}_{j\ell}$ and $\phi^{\rm S}_{j\ell}$
as the complex-range Gaussian basis.

The complex-range Gaussian basis functions
are oscillating with $r$.
They are therefore expected to simulate
the oscillating pattern of the continuous breakup state wave functions
better than the real-range Gaussian basis functions do.
Moreover, numerical calculation with the complex-range Gaussians
can be done using essentially the same computer programs as for the
real-range Gaussians, just replacing real variables for $a_{j}$
of Eq.~(\ref{aj}) by complex ones.
Usefulness of the real- and complex-range Gaussian basis
functions in few-body calculations are extensively presented
in the review work~\cite{GEM}.

%
\begin{table}
\caption{\baselineskip=6mm Test of  the 
accuracy of real-range and complex-range Gaussian
basis functions
for highly excited states 
($2n+l \leq 46,\: l=0 )$ of a harmonic 
oscillator potential for a nucleon. 
The number of basis functions is 28
for both cases.
Eigenenergies obtained by the diagonalization of the Hamiltonian
with the bases are listed in terms of the number of quanta,
$E/\hbar\omega-\frac{3}{2}$. See text for the Gaussian parameters.
}
\begin{tabular}{cccccccc}
\hline 
  &{\bf Exact }  & {\bf real}  
& \quad{\bf complex} & \qquad \qquad \qquad
  &{\bf Exact  }  & {\bf real}  & \quad{\bf complex}\quad  \\
  &{\bf (2n)}  & {\bf range}  
& \quad{\bf range} & \qquad \qquad \qquad
  &{\bf  (2n) }  & {\bf range}  & \quad{\bf range}\quad  \\
\hline 
\vspace{-3 mm} \\
&$\;0.0$   & $0.0000$ & $\:0.0000$ &   & 26.0 & 26.4 & 26.0001  \\
&$\;6.0$   & $\:6.0000$ & $\:6.0000$ &   & 30.0 & 32.9 & 30.0003  \\
&10.0   & 10.0000 & 10.0000 &   &  34.0 & 41.8 & $34.002\;\,$  \\
&14.0   & 14.0000 & 14.0000 &   &  38.0 & 53.8 & $38.003\;\,$  \\
&18.0   & $17.998\;\,$ & 18.0000 & &  42.0 & 69.9 & $42.1\;\;\;\;\;$ \\
&22.0 & $21.9\;\;\;\;\;$ & 22.0000 && 46.0 & 91.6 & $46.3\;\;\;\;\;$ \\
\hline
\end{tabular}
\label{table:accuracy}
\end{table}

\begin{figure}[htb]
\epsfig{file=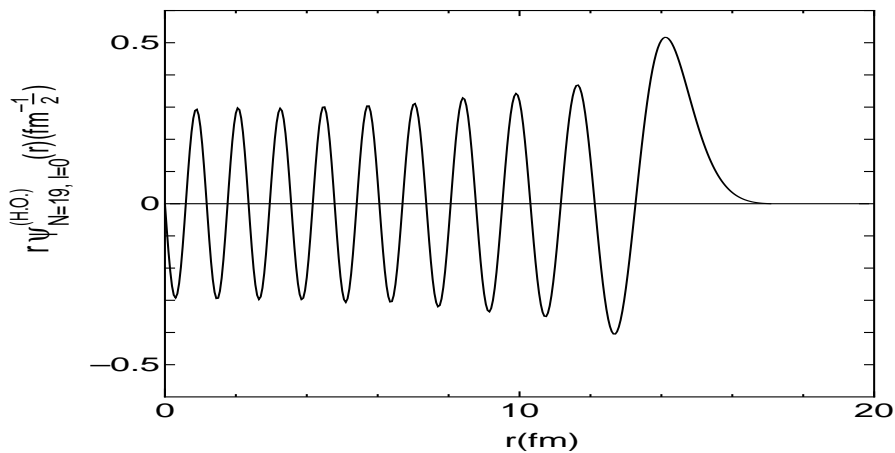,width=12cm,height=6cm}
\caption{\baselineskip=6mm 
Accuracy of the wave function of the  $2n+l=38,\, l=0 $ state 
obtained by diagonalizing Hamiltonian with a
harmonic-oscillator potential for a nucleon
using 28 complex-range Gaussian basis functions.
It is compared with the exact wave 
function but the difference is
invisible  since the error is less than a few \%
everywhere. See text for the Gaussian parameters.
This figure is taken from Ref.~\cite{GEM}.
}
\label{fig:howf}
\end{figure}

Here, we explore a typical example in which 
the complex-range Gaussian basis functions 
reproduce highly oscillatory functions with high accuracy.
A good test 
is to calculate the wave functions of 
highly excited states in a harmonic
oscillator potential; note that this potential
is not specially advantageous for the Gaussian bases. 
We take the case of a nucleon 
with  angular momentum $l=0$ in
a potential having $\hbar \omega=15.0 $ MeV.
Parameters of the complex-range
Gaussian basis functions are 
$\{\,2 n =28 , a_1=1.4 \, {\rm fm}, 
\,a_n=5.8 \,{\rm fm},
\: b=\frac{\pi}{2}\frac{1}{1.2^2}=1.09 \,\} $.
For the sake of comparison,
we also tested
the Gaussian basis functions with the parameters
$ \{ n=28,\:
a_1=0.5 \,{\rm fm},\: a_n=11.3\, $fm \}.
Optimized $a_1$ and $a_n$ are quite 
different between the two types of bases though the 
total numbers of basis functions  are the same.
In Table 1, 
we compare the calculated energy eigenvalues 
with the exact ones.  
It is evident that the complex-range Gaussians
can reproduce the energy up to much more highly excited states 
than the Gausssians do.  
For the Gaussian basis, even if the
number of basis functions is increased, 
the result is not significantly improved,
because the number of oscillation does not
increase. 
On the other hands,
for the complex-range Gaussian functions,
as  the number is increased, 
the result becomes better so long as the number of
oscillation is not too larger than $\sim$ 20.
Figure \ref{fig:howf} demonstrates 
good accuracy of the wave function
of the 19-th excited 
state having 38 quanta. Error is within a few \%,
much smaller than the thickness of the line.
The figure suggests that the basis functions 
is also suitable for describing 
pseudo-states used for Coulomb breakup reactions.

We here emphasize that even in the case where the projectile is
assumed to be three-body system, the Gaussian basis functions
with real and complex ranges 
are easily utilized in the CDCC calculation with the PS method.
We discuss this point in the next section.

Another  advantage of the PS method, 
in the case of two-body projectiles, is that 
the discrete breakup $S$-matrix elements, say $S_{nIL,0I_0 L_0}$,
for the transition from $\Phi_{0 I_0}(\xi)$
to $\Phi_{nI}(\xi)$ 
can be accurately transformed to smooth 
$S$-matrix elements, say $\tilde{S}_{IL,I_0 L_0}(k)$, 
as following \cite{ps-cdcc},
since the two-body PS basis functions can form in the good
approximation a complete set  in the finite region 
which is important for the breakup processes:
\begin{equation}
   \tilde{S}_{IL,0 I_0L_0}(k) = \sum_n  
\; \langle\tilde{\Phi}_{I}(k,\xi)|\Phi_{nI}(\xi)\rangle_\xi \;
    S_{nIL,0I_0L_0} \; ,
\label{S-matrix-appro}
\end{equation}
where $\tilde{\Phi}_{I}(k,\xi)$
is the exact wave function of the internal motion of the
two-body projectile.

\vskip 0.3cm
\noindent
{\bf  Example 1 $\: : \:^6$Li+$^{40}$C scattering at 156 MeV.
}

\begin{figure}[htbp]
  \includegraphics[width=0.5\textwidth,clip]{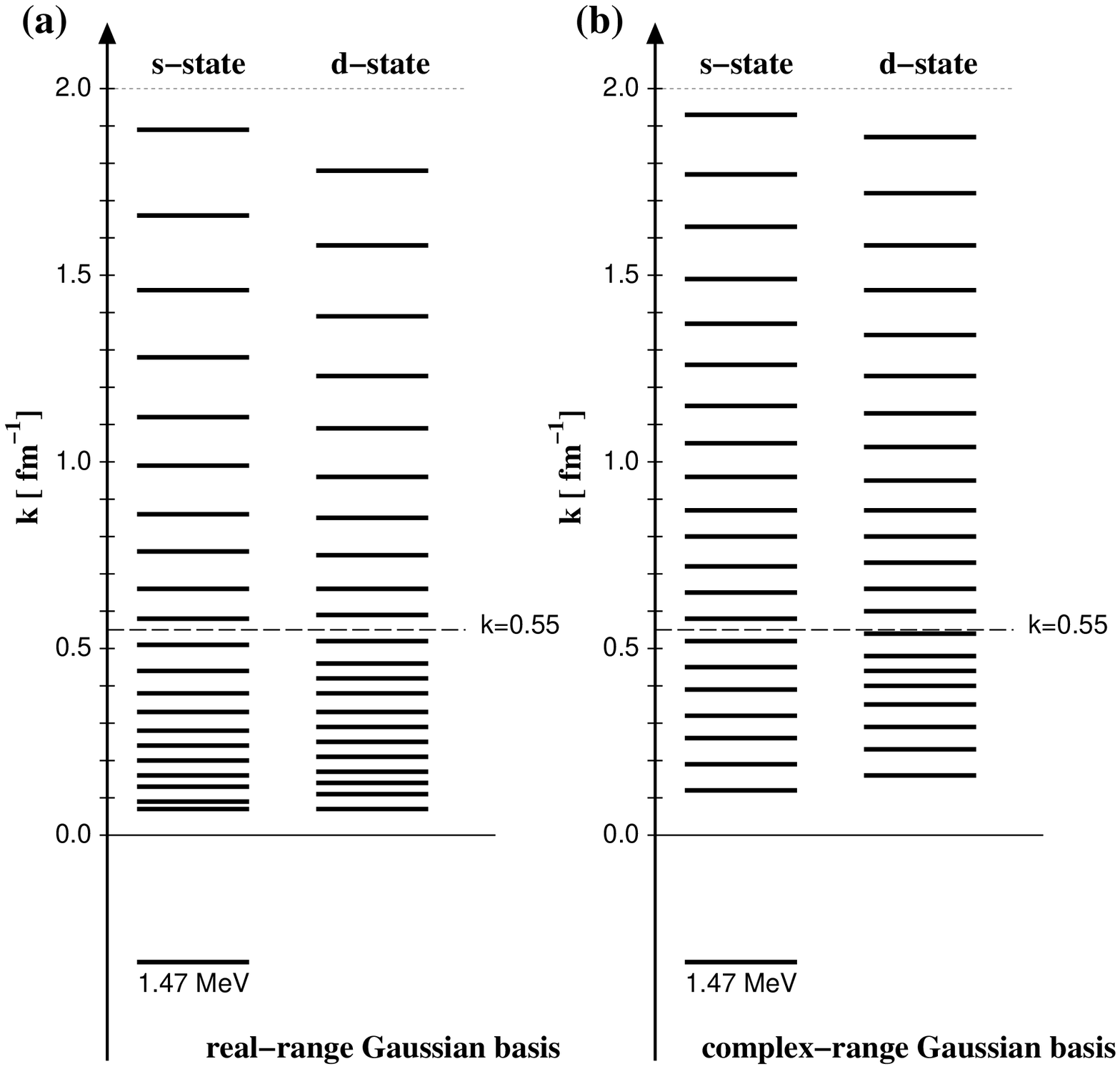}
 \caption{\baselineskip=6mm 
  Pseudo states (discretized continuum states) 
  for $^6$Li obtained 
  by using the real-range Gaussian basis functions
  (left) and the complex-range Gaussian basis functions
  (right).  This figure is taken from Ref.\cite{ps-cdcc}.
 }
\label{fig:Li6}
  \includegraphics[width=0.35\textwidth,clip]{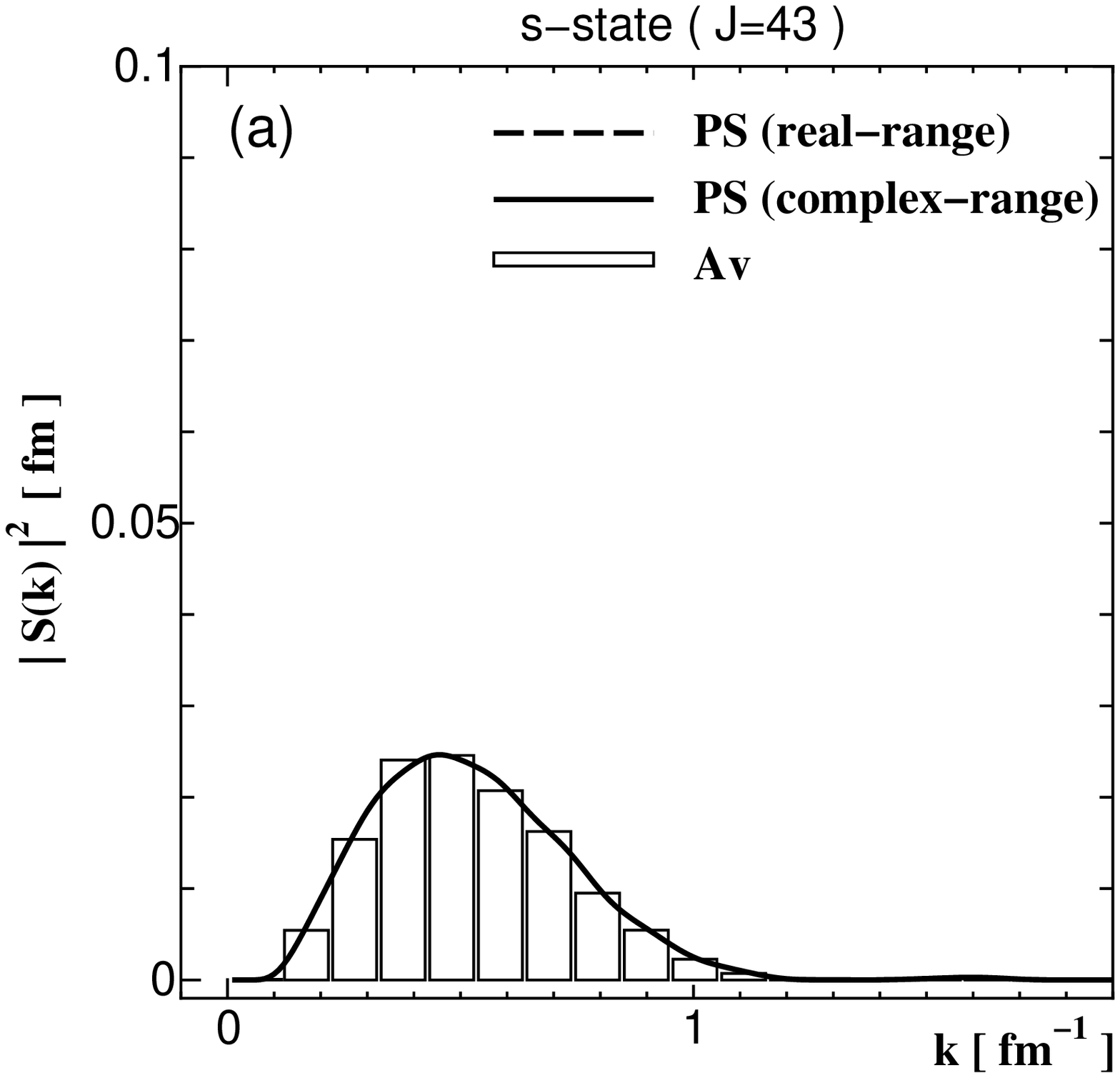}
  \includegraphics[width=0.35\textwidth,clip]{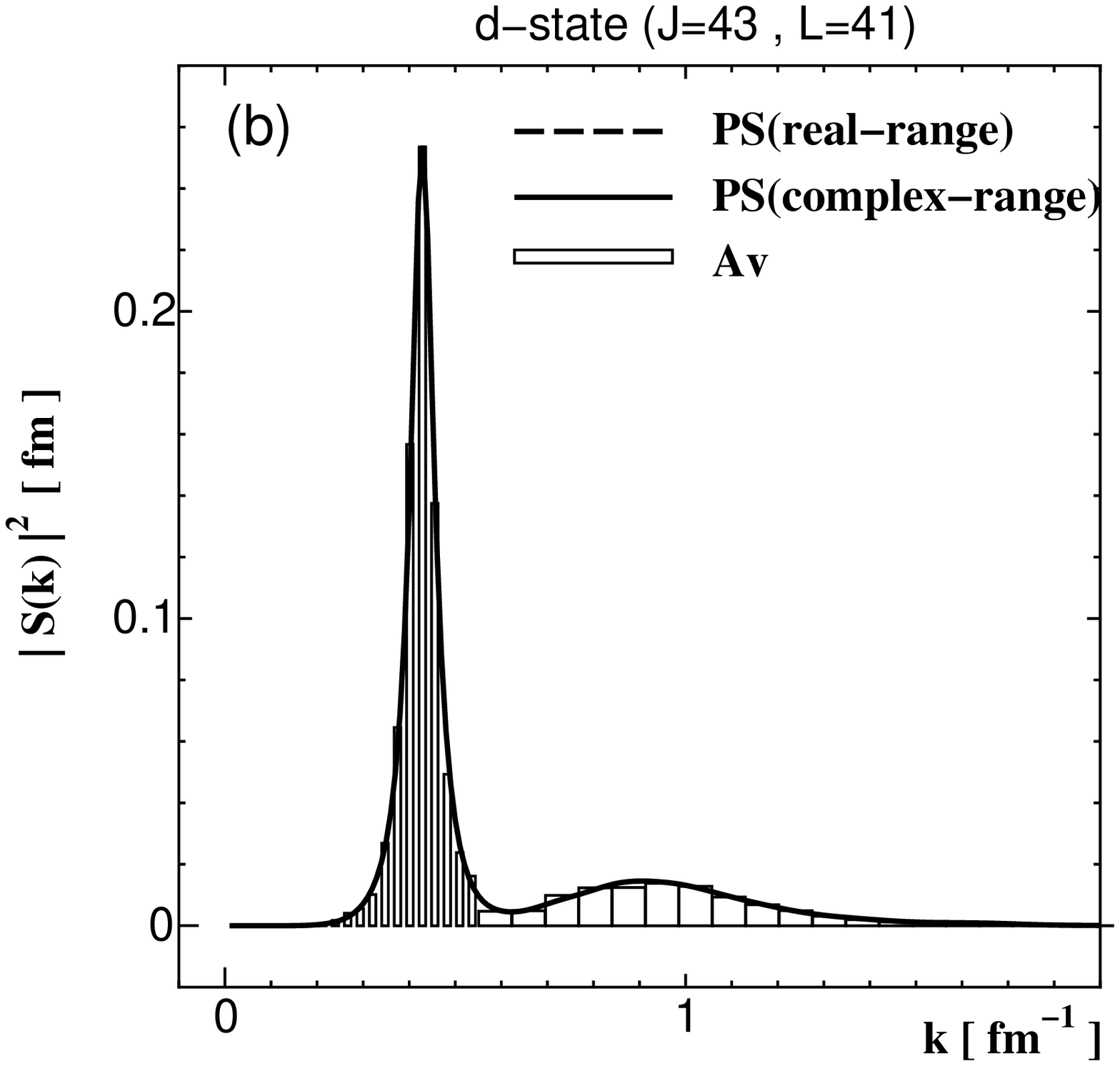}
  \caption{\baselineskip=6mm 
  The squared moduli of breakup $S$-matrix elements as a
  function of $k$ at the grazing total 
  angular momentum $J=43$ 
  for $^6$Li + $^{40}$Ca scattering   at 156 MeV. 
  The step line is the result of the momentum-bin
   method (Average-method) with dense bins. 
   $s$-state breakup (left) and $d$-state breakup
   for $L=J-2$ (right).    Note that
   the difference between the results of the real- and complex-range
   Gaussian PS methods is not visible since it is less than
   about 1\%.  This figure is taken from Ref.\cite{ps-cdcc}.
   } 
 \label{fig:Li6-S}
\end{figure}

Here, we briefly show results of test calculations done
in \cite{ps-cdcc}  $^6$Li+$^{40}$C scattering at 156 MeV.
The $\alpha-d$ continuum
of the $^6$Li projectile is 
discretized as in Fig.~\ref{fig:Li6}
using the real-range Gaussian bases and 
the complex-range Gaussian bases. 
The modelspace sufficient for describing breakup processes in this
scattering
is $k_{\rm max}=2.0$ fm$^{-1}$ and $\ell_{\rm max}=2$;
the modelspace is composed of two $k$-continua 
for $s$-state and $d$-state.
There exists a $d$-state resonance.
The resonance is automatically taken care by the 
PS method by the lowest-lying several pseudo-states. 
On the other hand, in the momentum-bin method,
the d-state $k$-continuum is further divided in the
momentum-bin method into
the resonant part $[0 < k < 0.55\, {\rm fm}^{-1}]$ 
and the non-resonant part
$[0.55 < k <2.0\, {\rm fm}^{-1}]$.
In the former region the $k$ continuum $d$-state wave function
varies rapidly with $k$.
The momentum-bin method can simulate this rapid change
with bins of an extremely small width.
In fact clear convergence is found for both the elastic and the breakup
$S$-matrix elements, when the resonance part is described by 30 bins and
the non-resonance part of the $d$-state and the $s$-state $k$-continua
by 20 bins.

Figure~\ref{fig:Li6-S} represents breakup $S$-matrix elements
at grazing total angular momentum $J=43$;
(a) $s$-state breakup and (b) $d$-state breakup in the case of
$L=J-2$.
The real- and complex-range Gaussian PS 
discretization well reproduce
the "exact" solution calculated by the 
momentum-bin method with dense bins.
The results of the two PS methods 
turn out to coincide within the
thickness of the line.
The resonance peak can be expressed by only 8 (12) breakup
channels in the complex-range (real-range) Gaussian PS method,
while the corresponding number of breakup channels 
is 30 in the momentum-bin method, as mentioned above.
Thus, one can conclude that the real- and complex-range Gaussian PS
methods are very useful for describing not only non-resonant states but
also resonant ones.

The PS method
has at least two advantages over the
widely used momentum bin average method.
One is that it does not need the
exact wave function of the projectile
over the entire region of $r$. This is important from a
theoretical point of view.
The other is that with the real- and complex-range
Gaussian bases one can calculate all the coupling potentials
semi-analytically~\cite{GEM},
which is very useful in actual calculations;
note that the Gaussian bases are very suitable for
transforming wave functions and interactions
from a Jacobian coordinate system to other ones.
Furthermore, if the projectile  has resonances
in its excitation spectrum,
the PS method discretizes the complicated spectrum
with a reasonable number of the pseudo-states,
without distinguishing the resonance states
from non-resonant continuous states.
These advantages of the PS method are
extremely helpful, sometimes even
essential, in applying CDCC to four-body breakup effects of
unstable nuclei such as $^6$He and $^{11}$Li.

\newpage
\noindent
{\bf Example 2 $\: : \:^8$B+$^{58}$Ni scattering at 25.8 MeV.
}
\begin{figure}[b]
 \begin{center}
  \includegraphics[width=0.35\textwidth,clip]{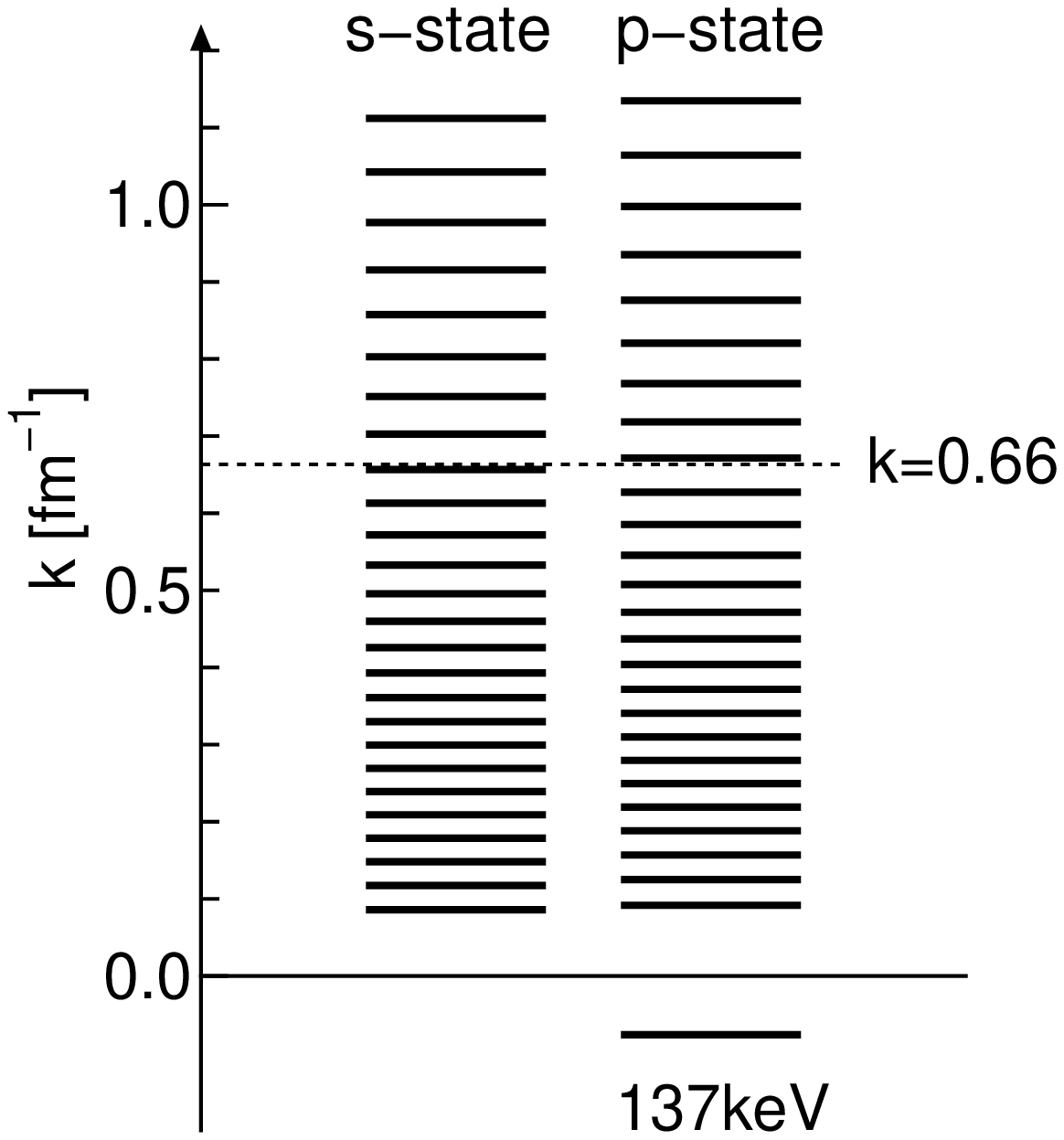}
\caption{\baselineskip=6mm Discretized momenta for $^8$B; the left (right) side corresponds
   to the s-state (p-state).
   The horizontal dotted line represents the cutoff
   momentum $k_{\rm max}$ taken to be 0.66 fm$^{-1}$
   above which is not effective in the reaction.
  This figure is taken from Ref.\cite{psc-cdcc}.
}
\label{fig:B8level}
\vskip 1.0cm
  \includegraphics[width=0.45\textwidth,clip]{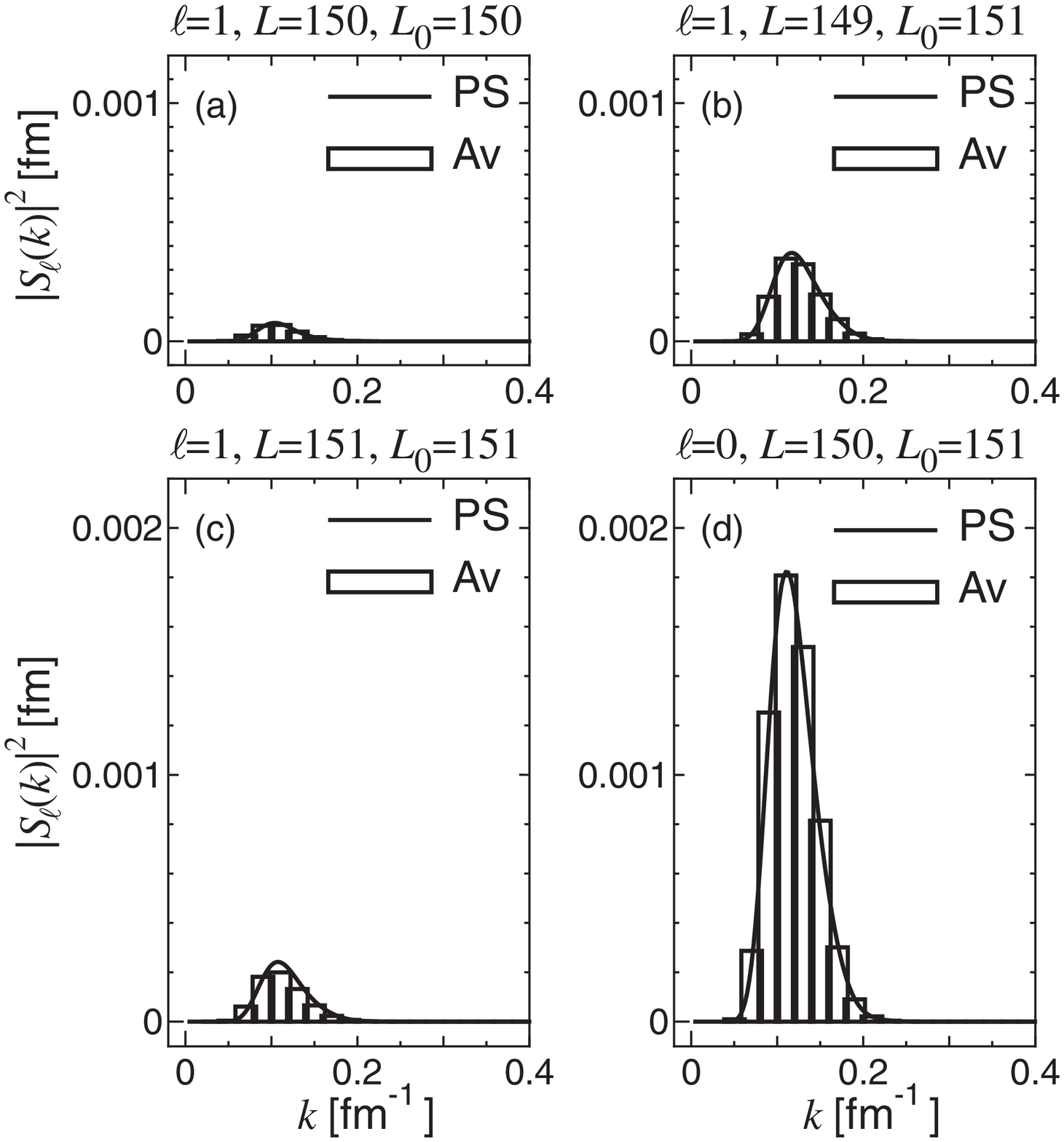}
  \caption{\baselineskip=6mm The squared moduli of breakup $S$-matrix elements,
  as a function of $k$, at $J=150$ for $^8$B+$^{58}$Ni
  scattering at 25.8 MeV.
  The panel (a), (b), (c) and (d) correspond to
  $(\ell,L,L_0)=(1,150,150)$, (1,149,151),
  (1,151,151) and (0,150,151), respectively.
  In each panel, the solid line represents the result of PS-CDCC,
  while the step line is the result 
  of the momentum-bin (average) method of 
  CDCC that is assumed
  as the {\lq\lq}exact'' $S$-matrix elements.
  This figure is taken from Ref.\cite{psc-cdcc}.
  }
 \end{center}
\label{fig:B8smat}
\end{figure}

Here, we briefly show results of test 
calculation in \cite{psc-cdcc}
for Coulomb breakup process of 
$^8$B+$^{58}$Ni scattering at 25.8 MeV.
 The $^7$Be$-p$ continuum in the $^8$B
projectile is discretized as in Fig.~5
by the PS method with the real-range Gaussian bases and 
the complex-range Gaussian bases. 
In the PS method, 
the number of channels included in
the CDCC calculation, 
was 18 for both the $s$- and $p$-states
at $k < k_{\rm max} = 0.66$ fm$^{-1}$,
which give a 
satisfactory convergence of the result. 
The resulting wave functions
with positive eigenenergies turned out to oscillate up to
about 100 fm.
In the momentum-bin method, 
the modelspace with $k_{\rm max}=0.66$ fm$^{-1}$ and
$\Delta k=0.66/16$ (0.66/32) fm$^{-1}$ for $p$-state 
($s$-state)
gives convergence of the resulting total breakup cross section.
The maximum internal coordinate $r_{\rm max}$ was taken
to be 100 fm.

Figure 6 shows the result of the comparison of
$|S_\ell (k)|^2$ at $J=150$, which
corresponds to the scattering angle of 10$^\circ$ assuming
the classical path. It was found that CDCC calculation with only
Coulomb coupling potentials gives a peak at 10$^\circ$ in the
total breakup cross section. Thus, it can be assumed that 
Fig.~6 corresponds to the most-Coulomb-like breakup process;
in any case, the feature of the result was found to be
almost independent of $J$.
In each panel of Fig.~6, 
one sees that the result of PS-CDCC (solid line)
very well reproduces the "exact" solution (step line by
the momentum-bin method)
for all $k$ being significant for the $^8$B Coulomb
breakup.

\section{GAUSSIAN EXPANSION METHOD FOR FEW-BODY SYSTEMS}

In this section we briefly explain the Gaussian expansion method
(GEM) for few-body systems.  The method was proposed 
by Kamimura in 1988 \cite{Kamimura88}
for three-body systems
and was much developed by Hiyama 
using the infinitesimally-shifted Gaussian basis functions
even for four-body systems (reviewed in \cite{GEM}).

A good example to show the accuracy and usefulness
of the method is the determination of  
upper limit of the difference between
the masses of proton and antiproton, $m_p$ and $m_{\bar p}$,
respectively.
The first recommended upper limit of 
$|m_{\bar p}-m_p|/m_p$ by the Particle Data Group
listed in Particle Listings 2000 
\cite{Listing2000}
was $5 \times 10^{-7}$, which could be used for a test of 
$CPT$ invariance.
This number was extracted from a high-resolution 
laser experiment involving
metastable states of 
antiprotonic helium atom (He$^{2+}+e^-+{\bar p}$)
\cite{Torii99} 
by Kino {\it et al.} \cite{Kino99}
through  a theoretical 
analysis of the highly excited states of
the Coulomb three-body system
using GEM. The ratio was improved to 
$|m_{\bar p}-m_p|/m_p < 1 \times 10^{-8}$, as listed in the 
Particle Listings  2004, by later,
more extensive experiments
and additional calculations (cf. Ref.\cite{GEM}) 

\begin{figure}[htbp]

 \includegraphics[width=0.5\textwidth,clip]{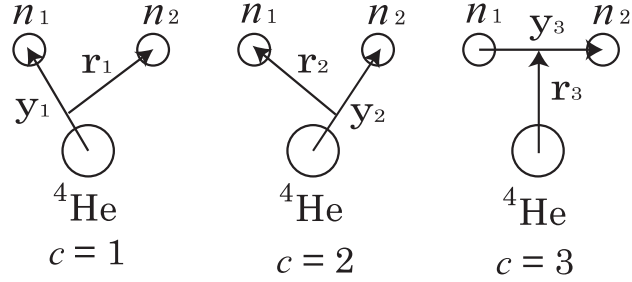}
 \caption{\baselineskip=6mm Jacobian coordinates of three rearrangement
 channels ($c=1\mbox{--}3$) adopted for the $n$+$n$+$^4$He model of $^6$He
 structure. The two neutrons are to be antisymmetrized.}
\label{fig:Jacobi}

\end{figure}

\begin{figure}[htbp]

 \includegraphics[width=0.85\textwidth,clip]{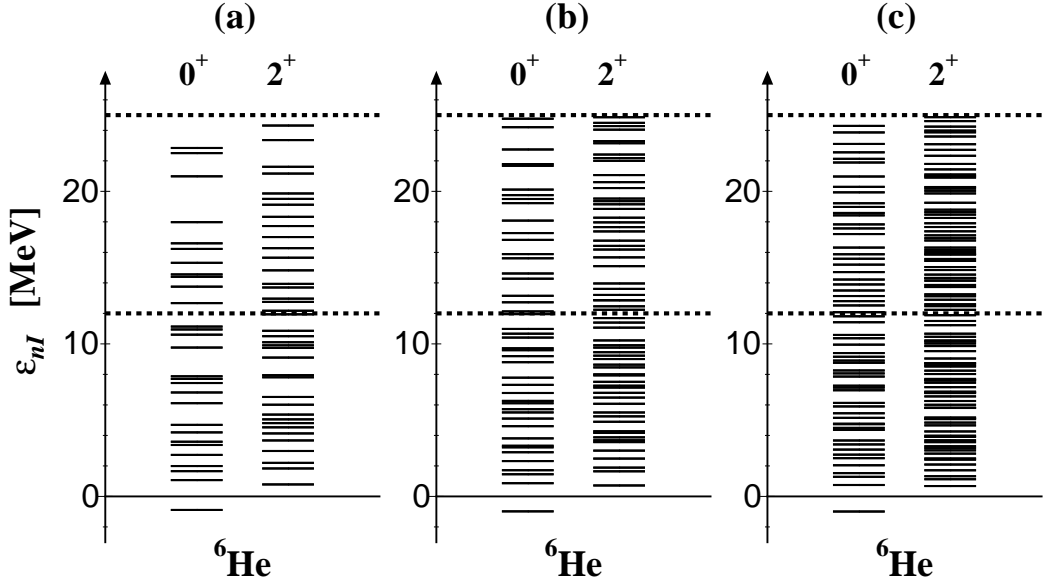}
 \caption{\baselineskip=6mm 
Calculated energy levels of the bound and discretized 
continuum states (pseudo-states) of $^6$He
using three different sets of Gaussian basis functions
(see Ref. \cite{fb-cdcc}). The numbers of the
pseudo-states used in the $^6$He+$^{12}$C scattering
at 229.8 MeV to see convergence of the calculated results 
is (a) 28 ($0^+$) and 39 ($2^+$),   
(b) 44 ($0^+$) and 64 ($2^+$) and
(c) 60 ($0^+$) and 85 ($2^+$) which are located in the 
region of 
$\epsilon_{nI} < 25$ MeV.  For 18-MeV scattering,
to take pseudo-states of $\epsilon_{nI} < 12$ MeV is
satisfactory.
The case (b) was found to
be sufficient to obtain a good convergence.
}
\label{fig:level}

\end{figure}

In the Gaussian expansion method~\cite{GEM},
wave functions of the projectile, 
$\Phi_{nIm}$ in (\ref{expansion}), is written as
a sum of component functions in 
the Jacobian coordinates for rearrangement channels
$c=1-3$ in Fig.~\ref{fig:Jacobi} as
\begin{equation}
\Phi_{nIm}(\xi)=\sum_{c=1}^3 \psi^{(c)}_{nIm}(\xi),
\label{phig}
\end{equation}
Each $\psi^{(c)}_{nIm}$ is expanded in terms of the Gaussian basis 
functions:
\begin{eqnarray}
\hskip -0.3cm \psi^{(c)}_{nIm}(\xi) &=&\varphi^{(\alpha)}
\sum_{\lambda \ell \Lambda S}
\sum_{i=1}^{i_{\rm max}}
\sum_{j=1}^{j_{\rm max}}
  A_{i\lambda j\ell \Lambda S}^{(c)nI} \,
\: y_c^{\lambda}\, r_c^{\ell} \,
  e^{-({y_c}/{\bar{y}_{i}})^2} \,
  e^{-({r_c}/{\bar{r}_{j}})^2} \nonumber \\
&& \hskip -0.3cm \times
   \Big[
   \left[
    Y_{\lambda}(\hat{\bf y}_c)
    \otimes Y_{\ell}(\hat{\bf r}_c)
  \right]_\Lambda
\otimes
\big[
\eta_{\frac{1}{2}}^{(n_1)}\otimes\eta_{\frac{1}{2}}^{(n_2)}
\big]_{S}
\Big]_{Im},
\label{gauex}
\end{eqnarray}
where $\lambda$ ($\ell$) is the angular momentum
regarding the Jacobian coordinates ${\bf y}_c$ (${\bf r}_c$),
and $\eta_{1/2}$ is the spin wave function of each valence neutron
($n_1$ or $n_2$).
$^4$He has been treated as an inert core
with the $(0s)^4$ internal configuration,
$\varphi^{(\alpha)}$.
The Gaussian range
parameters are taken to lie in geometric progression:
\begin{eqnarray}
 \bar{y}_{i}=\bar{y}_1
  (\bar{y}_{\rm max}/\bar{y}_1)^{(i-1)/(i_{\rm max}-1)},
\label{range1}
 \\
 \bar{r}_{j}=\bar{r}_1 (\bar{r}_{\rm max}/
  \bar{r}_1)^{(j-1)/(j_{\rm max}-1)}.
\label{range2}
\end{eqnarray}
$\Phi_{nIm}$ is
antisymmetrized for the exchange between $n_1$ and $n_2$.
Meanwhile, the exchange between each valence neutron and each nucleon
in $^4$He is treated approximately by the orthogonality condition.
The eigenenergies $\epsilon_{nI}$ of $^6$He
and the corresponding expansion-coefficients
$A_{i\lambda j\ell\Lambda S}^{(c)nI}$ are determined 
by diagonalizing the Hamiltonian
of the interrenal motion of $^6$He~\cite{GEM6He1,GEM6He}
using a large number of three-body Gaussian basis functions.
Detailed information on the basis  
is listed in Ref.\cite{fb-cdcc}.
The calculated $\epsilon_{nI}$ are $-0.98$ MeV for the $0^+$ ground
state and 0.72 MeV for the $2^+$ resonance state; 
here, we took the
Bonn A potential between the valence nucleons and 
increased the
depth of the $n-\alpha$ potential by a few percent
so that the ground-state energy is reproduced.

In the four-body CDCC calculation of $^6$He+$^{12}$C shown 
in a later section, we take 
$I^\pi=0^+$ and $2^+$ states for $^6$He. 
Here we omit the $1^-$  state 
that does not contribute to the
nuclear breakup processes (but they are included in the
calculation of Coulomb and nuclear breakup in Ref.\cite{fbc-cdcc}).
In order to demonstrate the convergence of the 
four-body CDCC solution
with increasing the number of the Gaussian basis functions,
we prepare three sets of the basis functions, i.e., sets I, II and
III listed in Table II of \cite{fb-cdcc}.
Resultant energy levels of the ground and pseudo-states
are shown in (a), (b) and (c) in Fig.~\ref{fig:level}, 
respectively.
For $^6$He+$^{12}$C
scattering at 18 MeV (229.8 MeV) which will be discussed
in the next section, high-lying states with
$\epsilon_{nI} > 12$ MeV ($\epsilon_{nI} > 25$ MeV) are found to
give no effect on the elastic and breakup $S$-matrix elements.
Thus, the effective number of the eigenstates of $^6$He,
is reduced much for each of cases (a), (b), (c) as shown in
Fig.~\ref{fig:level}. 
The case (b) was found to
be sufficient to obtain a good convergence.
In the GEM, computation time to obtain the wave functions
of the bound and pseudo states is very short;
for example, all the wave functions of the states 
in Fig.~\ref{fig:level}(c)
is obtained in 10 minutes on FUJITSU VPP5000, 
a supercomputer.


It is to be noted that the bound and pseudo-states obtained
with the GEM calculations construct an approximate
complete sets for each $J (=0,1,2)$ in a finite region which 
is responsible for the reaction; this was examined 
by checking that those states (below 100 MeV)  satisfies 
99.9 \% of the energy-weighted cluster sum-rule limit for
monopole, dipole and quadrupole transitions.

\section{Four-body CDCC analysis of 
$^6$H\lowercase{e}+$^{12}$C scattering at 18 and 229.8 MeV} 

In this section, we briefly introduce the results obtained
in the work of Ref.\cite{fb-cdcc}.
We performed the four-body CDCC calculation for
$^6$He+$^{12}$C scattering at 18 and 229.8 MeV using the
wave functions of the bound state and the pseudo-states
of $^6$He obtained above. 

The real part of the 
CC potentials, say $V^J_{nIL,n'I'L'}(R)$,
was constructed by using the double-folding
model \cite{dfm}; the potentials were calculated by
folding the DDM3Y $NN$ interaction
into the transition densities
between the states $\Phi_{nI}(\xi)$ and $\Phi_{n'I'}(\xi)$
(cf. Ref.\cite{fb-cdcc} for details)
and the ground-state density of $^{12}$C~\cite{Kamimura12C}
.
The imaginary part was assumed, as usually done
\cite{cdcc},
to be given as (together with the real part)
\begin{equation}
  (N_R+iN_I)\: V_{nIL,n'I'L'}^{J}(R) ,
\end{equation}
where $N_R=1.0$ with no renormarization of the real part.
The only parameter $N_I$ is searched for to reproduce
the observed elastic cross section as well as possible.
In the analysis of the 
$^6$He+$^{12}$C scattering, Coulomb breakup 
effect is ignored since it is negligible  
for this light target; the Coulomb potential is assumed to
work between the center-of-mass of the target and
that of the projectile.

\begin{figure}[htbp]
\includegraphics[width=0.45\textwidth,clip]{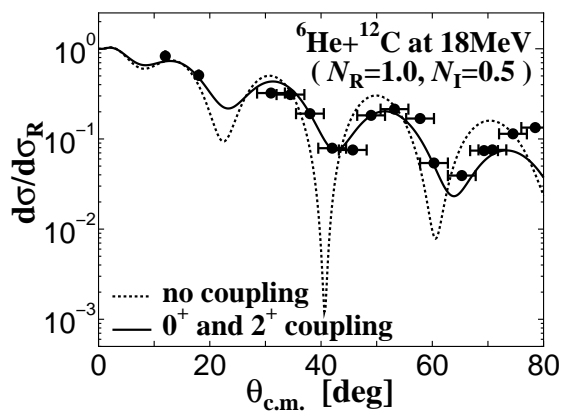}
\caption{\baselineskip=6mm Angular distribution of the elastic differential cross section
 for $^6$He+$^{12}$C scattering at
 18 MeV. The solid and dotted lines show the results with and
 without breakup effects, respectively. The experimental data are
 taken from Ref.~\cite{Milin}. This figure is taken from
 \cite{fb-cdcc}.
 }
\label{fig:elastic1}
\end{figure}

\begin{figure}[htbp]
\includegraphics[width=0.45\textwidth,clip]{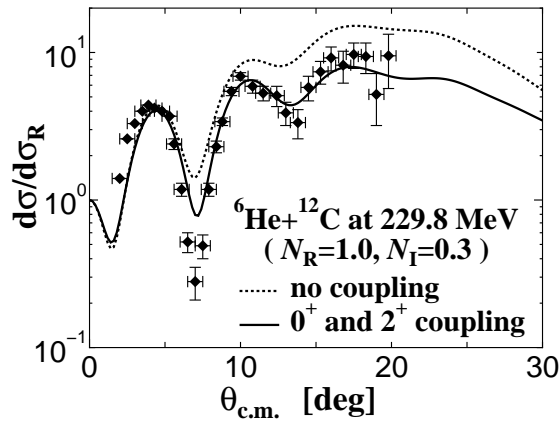}
\caption{\baselineskip=6mm The same as in Fig.~\ref{fig:elastic1} 
but for $^6$He+$^{12}$C
 scattering at 229.8 MeV. The experimental data are taken from
 Ref.~\cite{exp6He1}.
 This figure is taken from
 \cite{fb-cdcc}.}
\label{fig:elastic2}
\end{figure}

Calculated and observed elastic cross sections for
$^6$He+$^{12}$C scattering at 18  MeV 
are shown in Fig.~\ref{fig:elastic1}.
The optimum value of $N_I$ is 0.5, which  
is the same as that for $^6$Li scattering at various
incident energies \cite{cdcc}.
The dotted lines represent the elastic
cross sections due to the single-channel calculation. 
Then, the difference
between the solid and dotted lines shows the effect of
the four-body breakup on the elastic cross section.
The effect is sizable and indispensable to explain 
the behavior of the angular distribution.
The case at 229.8 MeV is shown in Fig.\ref{fig:elastic2}
and the optimum value of $N_I$ is 0.3.
The breakup effect in this case is also important
in reproducing the data. 
The origin of the small
$N_I$ value for the $^6$He scattering at 229.8 MeV
is not clear at this moment,
so more systematic
experimental data are highly desirable for $^6$He scattering.

We calculated the dynamical polarization (DP)
potential induced by
the four-body breakup processes, in order to understand effects of the
processes on the elastic scattering. 
The DP potential  is given by
the deviation of
 the so-called
wave-function-equivalent local potential derived using the
elastic channel amplitude 
in the solution of the CDCC equation
from the double-folding potential
of the elastic channel. 
From the analysis \cite{fb-cdcc} of the DP potential,
one sees that inclusion of the four-body breakup processes
makes the real part of the $^6$He--$^{12}$C potential 
shallower and the imaginary one deeper compared with 
the double-folding potential
of the elastic channel. 
In particular, the latter effect is 
important and can be assumed
to come from the Borromean structure of
$^6$He. This is consistent with the fact 
that the total reaction cross
section is enhanced by the Borromean structure\cite{fb-cdcc}.


\section{conclusion and near-future problems}

In conclusion, a fully quantum-mechanical method of treating 
four-body  breakup is presented by extending CDCC. 
The validity of the method called 
four-body CDCC is confirmed by clear convergence of
the calculated elastic and energy-integrated breakup 
cross sections with respect to extending the modelspace.
The four-body CDCC is found to explain well 
the  $^6$He+$^{12}$C 
scattering at 18 and 229.8 MeV in which $^6$He easily breaks up 
into two neutrons and $^4$He.
For the elastic scattering, the four-body breakup processes
make, in particular, the imaginary part of the $^6$He--$^{12}$C potential
deeper, which is originated in the Borromean structure of $^6$He.

In the analysis of \cite{fb-cdcc}, 
four-body Coulomb breakup is neglected.
However, it is possible to 
treat it within the four-body CDCC framework 
(cf. Fig.~\ref{fig:cdcc-th}).
Actually, after this RIA workshop,
we reported in Ref.\cite{fbc-cdcc} our
four-body CDCC calculation of the $^6$He+$^{209}$Bi scattering
at 19.0 and 22.5 MeV  taking both the Coulomb
and nuclear breakup effects into account.
The elastic cross sections were well reproduced by the
calculation.  So, the same framework will be applicable to
other cases of three-body projectiles with 
Coulomb and nuclear breakup.

In order to treat both Coulomb and nuclear breakup processes
at {\it intermediate energies}, Ref.~\cite{ecdcc} proposed
a new method, namely a hybrid calculation with
the three-body CDCC method and the eikonal-CDCC (E-CDCC) 
method. 
This hybrid calculation is expected to be 
opening the door to the systematic
analysis of Coulomb (plus nuclear) dissociation
of projectiles in the wide range of beam energies.
For example, the method was recently applied to 
the analysis of $^8$B dissociation measurements
to determine the astrophysical factor $S_{17}(0)$ accurately
\cite{astro}.

There are some important unstable nuclei
that are considered to be composed of 
four-body constituents. For reactions
in which such a four-body nucleus is a
projectile, a five-body CDCC calculation is required.
The GEM was already severely and successfully tested for 
the bound states and pseudo-states of four-body systems.
A good example is seen in a calculation of 
four-nucleon system ($^4$He) in Ref.~\cite{Hiyama04}.
The four-body GEM calculation with a realistic
$NN$ force (AV8') and a phenomenological $NNN$ force
(which is adjusted to reproduce the ground-state energy)
reproduced the energy of the second $0^+$ state 
and the $^4{\rm He}(e,e') ^4{\rm He}(0^+_2)$ form factor. 
Furthermore, some 3000 $\, 0^+$ pseudo-states below 300-MeV
excitation satisfied the energy-weighted monopole sum rule
by 99.9\% (with saturation) 
and made clear, for the first time, that
the major part of the monopole sum rule limit, which had
been long unknown, was distributed into low-lying
four-body non-resonant continuum states.
So, it may be said that it is ready to perform five-body CDCC
calculations for reactions induced by four-body projectiles.
 

\end{document}